\documentclass[english]{sbrt}
\usepackage[english]{babel}
\usepackage[utf8]{inputenc}

\usepackage[utf8]{inputenc}
\usepackage[english]{babel}
\usepackage[T1]{fontenc}
\usepackage{array}% para alinhamento de células na tabela
\usepackage{amsmath}
\usepackage{amsfonts}
\usepackage{amssymb}
\usepackage{amstext}

\usepackage{amsthm}
\usepackage{mathrsfs}
\usepackage{color}
\usepackage{listings}
\usepackage{placeins}
\usepackage{graphicx}
\usepackage[colorlinks=true,citecolor=blue,linkcolor=blue,urlcolor=blue]{hyperref}
\usepackage{mathtools}
\usepackage{times}
\usepackage{latexsym}
\usepackage[running,mathlines]{lineno}
\usepackage{verbatim}
\usepackage{bm}
\usepackage{ulem}
\usepackage{tcolorbox}
\usepackage{orcidlink}

\usepackage{xcolor}  % Para definir cores personalizadas
\usepackage{ragged2e}

\usepackage{lipsum}
\usepackage[activate={true,nocompatibility},final,tracking=true,kerning=true,spacing=true,factor=1100,stretch=10,shrink=10]{microtype}
\tcbuselibrary{minted,breakable,skins,xparse}
\definecolor{bg}{gray}{0.95}

\definecolor{codegreen}{rgb}{0,0.6,0}
\definecolor{codegray}{rgb}{0.5,0.5,0.5}
\definecolor{codepurple}{rgb}{0.58,0,0.82}
\definecolor{backcolour}{rgb}{0.95,0.95,0.92}

\lstdefinestyle{mystyle}{
    backgroundcolor=\color{backcolour},   
    commentstyle=\color{codegreen},
    keywordstyle=\color{magenta},
    numberstyle=\tiny\color{codegray},
    stringstyle=\color{codepurple},
    basicstyle=\ttfamily\footnotesize,
    breakatwhitespace=false,         
    breaklines=true,                 
    captionpos=b,                    
    keepspaces=true,                 
    numbers=left,                    
    numbersep=5pt,                  
    showspaces=false,                
    showstringspaces=false,
    showtabs=false,                  
    tabsize=2
}

\newtheoremstyle{mytheoremstyle} % name of the style to be used
  {0.5cm}                        % measure of space to leave above the theorem. E.g.: 0.5cm
  {0.5cm}                        % measure of space to leave below the theorem. E.g.: 0.5cm
  {}                     % name of font to use in the body of the theorem
  {}                             % measure of space to indent
  {\bfseries}                    % name of head font
  {.}                            % punctuation between head and body
  {0.5em}                        % space after theorem head
  {}                             % Manually specify head

\theoremstyle{mytheoremstyle}
\theoremstyle{mytheoremstyle}
\theoremstyle{mytheoremstyle}

\begin{document}

\title{Fixed Point Exploration For CV-QKD IR QC-MET-LDPC Toward Hardware Implementation}

\author{Guilherme Vergne de Oliveira\orcidlink{0009-0008-0192-7648}, Mauro Queiroz Nooblath Neto\orcidlink{0000-0003-2972-9463}, Micael Andrade Dias\orcidlink{0000-0001-6394-9174}, Francisco Revson Fernandes Pereira\orcidlink{0000-0001-5638-6334}, Francisco Marcos de Assis\orcidlink{0000-0002-8619-0874}, Valéria Loureiro da Silva\orcidlink{0000-0001-5466-7933} and Nelson Alves Ferreira Neto\orcidlink{0000-0003-2278-1082}
\thanks{Guilherme Vergne de Oliveira\orcidlink{0009-0008-0192-7648}, QuIIN – Quantum Industrial Innovation, EMBRAPII CIMATEC Competence Center in Quantum Technologies, SENAI CIMATEC, Salvador-Bahia, e-mail: guilhermevergne@gmail.com; Mauro Queiroz Nooblath Neto\orcidlink{0000-0003-2972-9463}, QuIIN – Quantum Industrial Innovation, EMBRAPII CIMATEC Competence Center in Quantum Technologies, SENAI CIMATEC, Salvador-Bahia, e-mail: mauro.neto@fieb.org.br; Micael Andrade Dias\orcidlink{0000-0001-6394-9174}, Department of Electrical and Photonics Engineering, Technical University of Denmark, Lyngby-Denmark, e-mail: mandi@dtu.dk; Francisco Revson Fernandes Pereira\orcidlink{0000-0001-5638-6334}, IQM Quantum Computers, Georg-Brauchle-Ring 23–25, 80992 Munich, Germany, e-mail: revson.ee@gmail.com; Francisco Marcos de Assis\orcidlink{0000-0002-8619-0874}, Departamento de Engenharia Elétrica, UFCG - Universidade Federal de Campina Grande, Campina Grande-Paraíba, e-mail: fmarcos@dee.ufcg.edu.br; Valéria Loureiro da Silva\orcidlink{0000-0001-5466-7933}, QuIIN – Quantum Industrial Innovation, EMBRAPII CIMATEC Competence Center in Quantum Technologies, SENAI CIMATEC, Salvador-Bahia, e-mail: valeria.dasilva@fieb.org.br; Nelson Alves Ferreira Neto\orcidlink{0000-0003-2278-1082}, QuIIN – Quantum Industrial Innovation, EMBRAPII CIMATEC Competence Center in Quantum Technologies, SENAI CIMATEC, Salvador-Bahia, e-mail: nelson.neto@fieb.org.br.}
}

\maketitle

%\markboth{XLIV BRAZILIAN SYMPOSIUM ON TELECOMMUNICATIONS AND SIGNAL PROCESSING - SBrT 2026, SEPTEMBER 29TH TO OCTOBER 2ND, 2026, SALVADOR, BA}{}

\begin{abstract}
High-speed LDPC decoding is a major bottleneck in CV-QKD and motivates hardware acceleration with fixed-point arithmetic. This work compares SPA, MSA, and NMS under a unified low-SNR fixed-point framework using common graph, matrix, and quantization settings. Multiple formats are evaluated through FER, and average iterations. The results show that performance depends strongly on the interaction between decoder rule and numerical precision. SPA achieved the best overall performance. For reduced-complexity decoders, Q16.8 was the lowest consistent precision, with NMS outperforming MSA. Practically, SPA with Q8.4 offered the best balance between reliability and hardware efficiency for large-scale implementations.
\end{abstract}
\begin{keywords}
Fixed-point arithmetic, METLDPC codes, low-SNR decoding, hardware-oriented implementation.
\end{keywords}

\section{Introduction}

Low-density parity-check (LDPC) codes play a central role in modern error-correction systems operating close to channel capacity, especially in scenarios where reliable communication must be maintained at very low signal-to-noise ratio (SNR), typically around or below \(0\) dB and, in severe operating conditions, in the negative-SNR regime \cite{mani2021multiedge, wang2018high}. This regime is particularly relevant in reconciliation architectures for continuous-variable quantum key distribution (CV-QKD), where the legitimate users, conventionally referred to as Alice and Bob, must convert correlated continuous variables into a common binary key after transmission through a noisy quantum channel. In this context, decoding efficiency and error-correction capability directly impact the achievable secret key rate and transmission distance \cite{jouguet2011long, zhang2024review}. Structured low-rate constructions, typically associated with code rates far below those used in conventional communication systems and often below \(0.1\), such as multi-edge LDPC codes, have therefore received increasing attention due to their favorable balance between decoding performance and implementation feasibility \cite{Milicevic_2018, yang2020high}.

Although floating-point implementations are useful for algorithmic validation, practical high-throughput decoders intended for hardware platforms such as FPGAs must operate under strict constraints on memory footprint, arithmetic complexity, latency, and power consumption \cite{yang2021fpga}. For this reason, fixed-point arithmetic becomes a natural requirement in the transition from software-based simulation to hardware-oriented realization. However, reducing numerical precision introduces non-negligible effects, including quantization noise, restricted dynamic range, saturation, and sensitivity to intermediate products and accumulations. These effects are especially critical in low-SNR decoding, where small differences in message reliability may determine whether iterative convergence succeeds or fails \cite{zhou2022highthroughputdecoderquasicyclicldpc,cil2024loglogdomainsumproductalgorithm,yang2021fpga}.

The literature already reports relevant advances on limited-precision LDPC decoding, both from the perspective of hardware implementation and from the standpoint of low-bit-width message-passing strategies \cite{zhou2022highthroughputdecoderquasicyclicldpc,wang2022reconstructioncomputationquantizationrcqparadigmlow,geiselhart2022learningquantizationldpcdecoders,chen2005reduced}. In parallel, multi-edge constructions have been investigated as promising candidate for reconciliation in long-distance CV-QKD systems \cite{Milicevic_2018, zhou2021ratecompatible}. Nevertheless, a direct and controlled comparison between distinct decoding algorithms under the same fixed-point experimental framework remains relevant, particularly in the low-SNR regime. In practice, the numerical behavior of the sum-product algorithm (SPA), the min-sum algorithm (MSA), and the normalized min-sum algorithm (NMS) may differ substantially when subjected to the same clipping, scaling, and word-length restrictions \cite{ferraz2022survey, chen2005reduced}.

In this work, we present a unified fixed-point decoding framework for the comparative evaluation of SPA, MSA, and NMS. The three decoders are implemented under the same sparse graph representation, the same parity-check matrix, and the same quantization pipeline, allowing a fair comparison of the impact of arithmetic precision on decoding behavior. The analysis considers distinct fixed-point configurations, parameterized by total word length and fractional precision, and evaluates their effect on frame error rate (FER) and average number of iterations. In addition, the proposed framework makes it possible to discuss practical implementation effects, such as saturation and intermediate-product overflow.

\subsection{Multidimensional Reconciliation}

In CV-QKD, Alice and Bob initially share correlated Gaussian variables rather than binary sequences. In the low-SNR regime, direct reconciliation over these continuous observations becomes inefficient. Multidimensional reconciliation (MDR) addresses this issue by transforming the original Gaussian correlation into an equivalent binary-input additive white Gaussian noise (BI-AWGN) channel, enabling the use of binary error-correcting codes such as LDPC codes \cite{Leverrier2008Multidimensional}. In this work, MDR is not the main object of investigation, but the reconciliation strategy that motivates the decoding stage analyzed in the following sections.

\subsection{MET-LDPC}

Multi-edge type low-density parity-check (MET-LDPC) codes generalize irregular LDPC ensembles by describing the Tanner graph through multiple edge classes rather than a single aggregated degree distribution \cite{Milicevic_2018}. This allows variable and check nodes to be characterized not only by their total degree, but also by the distribution of edge types connected to them, providing finer structural control.

This flexibility is especially relevant in very low-SNR reconciliation scenarios, where conventional irregular degree profiles may be insufficient to sustain efficient iterative decoding. MET-LDPC ensembles are commonly described by multivariate degree-distribution polynomials, enabling more detailed graph design at the cost of increased construction complexity. In this work, MET-LDPC codes are relevant as a structural framework suitable for low-rate, low-SNR decoding conditions.

\subsection{Fixed-Point Modeling}

The transition from floating-point to fixed-point arithmetic is a fundamental step when moving from algorithmic validation to hardware-oriented decoder design. In practical implementations, especially in FPGA-based architectures, arithmetic precision must be explicitly constrained in order to reduce memory usage, simplify datapaths, and improve throughput. However, this reduction in numerical precision may also affect decoder convergence, message reliability, and overall error-correction performance, particularly in low-SNR regimes \cite{zhou2022highthroughputdecoderquasicyclicldpc,wang2022reconstructioncomputationquantizationrcqparadigmlow,cil2024loglogdomainsumproductalgorithm}.

In the fixed-point model adopted in this work, real-valued quantities are represented by signed integers associated with a scaling factor determined by the number of fractional bits. Let \(n_t\) denote the total word length and \(n_f\) the number of fractional bits. A real-valued quantity \(x \in \mathbb{R}\) is quantized as
\begin{equation}
x_q = \mathrm{round}\!\left(x\,2^{n_f}\right),
\label{eq:fp_quant}
\end{equation}
where \(x_q\) is the corresponding integer representation. The inverse mapping is approximately given by
\begin{equation}
x \approx \frac{x_q}{2^{n_f}}.
\label{eq:fp_dequant}
\end{equation}
Therefore, \(n_f\) controls the resolution of the representation, while \(n_t\) determines the available dynamic range.

For a signed fixed-point format with \(n_t\) total bits, the representable integer interval is
\begin{equation}
q_{\min} = -2^{\,n_t-1},
\qquad
q_{\max} = 2^{\,n_t-1}-1.
\label{eq:fp_limits}
\end{equation}
Any value exceeding this interval must be clipped or saturated to the nearest representable bound. In addition, before quantization, channel log-likelihood ratios (LLRs) are clipped to the interval
\begin{equation}
[-L_{\max},\,L_{\max}],
\label{eq:llr_clip}
\end{equation}
where \(L_{\max}\) is a design parameter that limits the dynamic range of the decoder messages. This clipping operation is particularly important in low-SNR decoding, since excessively large message magnitudes may produce numerical instability without providing significant reliability gains \cite{cil2024loglogdomainsumproductalgorithm}.

% To illustrate the adopted representation, consider a fixed-point format with \(n_t = 16\) and \(n_f = 8\). In this case, the scaling factor is \(2^8 = 256\). If the real-valued message is \(x = 1.5\), then its quantized form is
% \begin{equation}
% x_q = \mathrm{round}(1.5 \times 256) = 384.
% \end{equation}
% Applying \eqref{eq:fp_dequant}, the reconstructed value is
% \begin{equation}
% x \approx \frac{384}{256} = 1.5.
% \end{equation}
% Similarly, if \(x = 0.75\), then
% \begin{equation}
% x_q = \mathrm{round}(0.75 \times 256) = 192.
% \end{equation}
% A fixed-point multiplication between these two values is first carried out in the integer domain,
% \begin{equation}
% 384 \times 192 = 73728,
% \end{equation}
% and then rescaled by shifting \(n_f\) bits to the right:
% \begin{equation}
% \hat{x}_q = \mathrm{round}\!\left(\frac{73728}{256}\right) = 288.
% \end{equation}
% The corresponding real-valued result is
% \begin{equation}
% \hat{x} \approx \frac{288}{256} = 1.125,
% \end{equation}
% which matches the expected product \(1.5 \times 0.75 = 1.125\). This example illustrates the central idea of fixed-point arithmetic: operations are performed on scaled integers, and the scale is explicitly restored after multiplication.

In practice, the use of finite precision introduces two distinct numerical effects. The first is quantization itself, which limits the resolution with which message magnitudes can be represented. The second is saturation, which limits the maximum representable magnitude. In iterative LDPC decoding, both effects are relevant because messages are repeatedly updated, accumulated, and transformed along the Tanner graph. As a consequence, the chosen word length may affect not only FER, but also convergence speed and numerical stability \cite{wang2022reconstructioncomputationquantizationrcqparadigmlow,geiselhart2022learningquantizationldpcdecoders}.

Under the model adopted here, the three decoding algorithms considered in this paper share the same fixed-point infrastructure. Channel LLRs are clipped and quantized according to \eqref{eq:fp_quant}--\eqref{eq:llr_clip}, variable-node messages are accumulated under saturation constraints, and decoder outputs are kept within the representable range defined by \eqref{eq:fp_limits}. The main difference lies in the check-node update: MSA uses the standard min-sum rule, NMS applies a normalization factor \(\alpha\), and SPA employs nonlinear reliability updates while keeping message storage in fixed-point form. This unified representation makes it possible to isolate the influence of arithmetic precision from the influence of the decoding rule itself.

The fixed-point model is therefore not only a numerical approximation, but also an abstraction of the constraints expected in future hardware realization. By varying \(n_t\) and \(n_f\), one can investigate the trade-off between decoding performance and implementation cost, identifying precision regimes that preserve reliable operation while reducing arithmetic complexity \cite{zhou2022highthroughputdecoderquasicyclicldpc,wang2022reconstructioncomputationquantizationrcqparadigmlow}.

\subsection{Decoding Algorithms}

The decoding framework considered in this work is based on iterative message passing over the Tanner graph associated with the parity-check matrix. In all cases, reliability messages are exchanged between variable nodes and check nodes until either the target syndrome is satisfied or the maximum number of iterations is reached. Under the fixed-point model introduced in the previous subsection, the three algorithms analyzed here---SPA, MSA, and NMS---share the same message representation and the same variable-node update rule. Their main difference lies in the check-node processing. 

Let \(L_{j}^{\mathrm{ch}}\) denote the channel log-likelihood ratio (LLR) associated with variable node \(j\), \(L_{j \to i}\) the message sent from variable node \(j\) to check node \(i\), and \(L_{i \to j}\) the message sent from check node \(i\) to variable node \(j\). 
% \begin{equation}
% L_{j \to i} = L_{j}^{\mathrm{ch}} + \sum_{i' \in \mathcal{N}(j)\setminus i} L_{i' \to j},
% \label{eq:var_update}
% \end{equation}
% where \(\mathcal{N}(j)\) denotes the set of checks connected to variable node \(j\). The corresponding a posteriori reliability used for hard decision is
% \begin{equation}
% L_{j}^{\mathrm{app}} = L_{j}^{\mathrm{ch}} + \sum_{i \in \mathcal{N}(j)} L_{i \to j}.
% \label{eq:app_llr}
% \end{equation}
% Thus, the distinction between the decoding algorithms arises from how the check-to-variable messages \(L_{i \to j}\) are computed.

In the sum-product algorithm (SPA), the check-node update follows the standard belief-propagation rule in the LLR domain \cite{Moon2021}:
\begin{equation}
L_{i \to j}
=
2 \tanh^{-1}
\left(
\prod_{j' \in \mathcal{N}(i)\setminus j}
\tanh\left(\frac{L_{j' \to i}}{2}\right)
\right),
\label{eq:spa_check}
\end{equation}
where \(\mathcal{N}(i)\) denotes the set of variable nodes connected to check node \(i\). Among the algorithms considered in this work, SPA provides the most accurate reliability update, since it preserves the nonlinear form derived from the exact message-passing equations. For this reason, it is used here as the main decoding reference, although its implementation is also the most demanding from the numerical and architectural points of view.

The min-sum algorithm (MSA) replaces the nonlinear check-node rule by a reduced-complexity approximation based on the sign product and the minimum magnitude of the incoming messages \cite{Moon2021}:
\begin{equation}
L_{i \to j}
\approx
\left(
\prod_{j' \in \mathcal{N}(i)\setminus j}
\mathrm{sign}(L_{j' \to i})
\right)
\min_{j' \in \mathcal{N}(i)\setminus j}
\left|L_{j' \to i}\right|.
\label{eq:msa_check}
\end{equation}
This approximation substantially simplifies the check-node computation and makes MSA attractive in implementation-oriented scenarios. However, since it neglects the correction term implicitly present in the SPA formulation, MSA tends to overestimate message magnitudes, which may degrade decoding performance depending on the operating regime.

To reduce this discrepancy, the normalized min-sum algorithm (NMS) introduces a scaling factor \(\alpha \in (0,1]\) in the min-sum update \cite{1001666}:
\begin{equation}
L_{i \to j}^{\mathrm{NMS}}
=
\alpha
\left(
\prod_{j' \in \mathcal{N}(i)\setminus j}
\mathrm{sign}(L_{j' \to i})
\right)
\min_{j' \in \mathcal{N}(i)\setminus j}
\left|L_{j' \to i}\right|.
\label{eq:nms_check}
\end{equation}
When \(\alpha = 1\), NMS reduces to the conventional MSA. For \(\alpha < 1\), the outgoing check-node messages are attenuated, which generally leads to a better approximation of the SPA behavior while preserving the simplicity of the min-sum structure. As discussed by Chen and Fossorier \cite{1001666}, this normalization improves the convergence behavior of min-sum-based decoding and can yield a more favorable performance-complexity trade-off.

In the fixed-point framework adopted in this paper, the three algorithms are evaluated under the same sparse graph representation, the same quantization pipeline, and the same saturation constraints. This allows the comparison to focus on the combined effect of arithmetic precision and check-node update rule. In particular, SPA serves as a higher-complexity decoding reference, MSA represents the simplest approximation, and NMS occupies an intermediate position by introducing a controlled correction to the min-sum rule.

\section{Experimental Configuration}
\label{sec: sec2A}

The experimental evaluation was conducted under a common simulation framework in order to compare the fixed-point behavior of the SPA, MSA, and NMS decoders under identical decoding conditions. For all experiments, the same parity-check matrix, channel model, stopping criterion, and maximum number of iterations were adopted, so that the observed differences could be associated primarily with the decoding rule and the fixed-point precision.

Decoder performance was evaluated in terms of frame error rate (FER), and average number of iterations. Considering \(N_{\mathrm{frames}}\) simulated frames of length \(N\), the FER is defined as
\begin{equation}
\mathrm{FER} = \frac{N_{\mathrm{frame\_err}}}{N_{\mathrm{frames}}},
\end{equation}
where \(N_{\mathrm{frame\_err}}\) is the number of frames containing at least one residual bit error. In addition, as \(I_k\) denotes the number of iterations associated with frame \(k\), the average number of iterations is computed as
\begin{equation}
\bar{I} = \frac{1}{N_{\mathrm{frames}}}\sum_{k=1}^{N_{\mathrm{frames}}} I_k.
\end{equation}

The simulations were performed over a low-SNR range, with
\begin{equation}
\mathrm{SNR}_{\mathrm{dB}} \in \{-20.0,\,-19.5,\,\ldots,\,-15.0\},
\end{equation}
using a step of \(0.5\) dB. For each SNR point, \(2000\) frames were simulated. The codeword length was fixed at \(N=10000\), and the design rate was set to \(R=0.02\). The code ensemble was generated from the MET-LDPC polynomials \(\nu(x)=0.02x_1^2x_2^{51}+0.02x_1^3x_2^{60}+0.96x_3\) and \(\mu(x)=0.016x_1^4+0.004x_1^9+0.30x_2^3x_3+0.66x_2^2x_3\), with normalized check-node fractions, being \(\nu(x)\) the variable polynomial and \(\mu(x)\) the check polynomial..
% \begin{equation}
%     \nu(x)=0.02\,x_1^2 x_2^{51}+0.02\,x_1^3 x_2^{60}+0.96\,x_3
% \end{equation}
% \begin{equation}
%     \mu(x)=0.016x_1^4+0.004x_1^9+0.30x_2^3x_3+0.66x_2^2x_3
% \end{equation}
% Being \(\nu(x)\) the variable polynomial and \(\mu(x)\) the check polynomial.

The fixed-point analysis was carried out by varying the total word length and the number of fractional bits. The tested configurations were
\begin{equation}
Q8.4,\quad Q12.6,\quad Q16.8,\quad Q24.12,
\end{equation}
all evaluated with the same clipping limit \(L_{\max}=8.0\) and the same maximum number of decoding iterations, \(I_{\max}=100\). In this notation, \(Qn_t.n_f\) denotes a signed fixed-point representation with \(n_t\) total bits and \(n_f\) fractional bits.

For NMS, \(\alpha\) was preliminarily swept from 0.40 to 0.85 in steps of 0.05, and \(\alpha=0.65\) was selected for the main experiments. The MSA results correspond to the particular case \(\alpha = 1\), while SPA was evaluated separately through its own check-node update rule.

To ensure a fair comparison, all decoders were tested using the same sparse graph representation, the same quantization pipeline for a given fixed-point configuration, and the same simulation procedure at each SNR point. In this way, the analysis isolates the effect of arithmetic precision and decoding rule, allowing the comparison to focus directly on the reliability and convergence behavior of SPA, MSA, and NMS in the low-SNR regime.

\section{Results and Discussion}

\begin{figure}[H]
    \centering
    \includegraphics[width=0.98\linewidth]{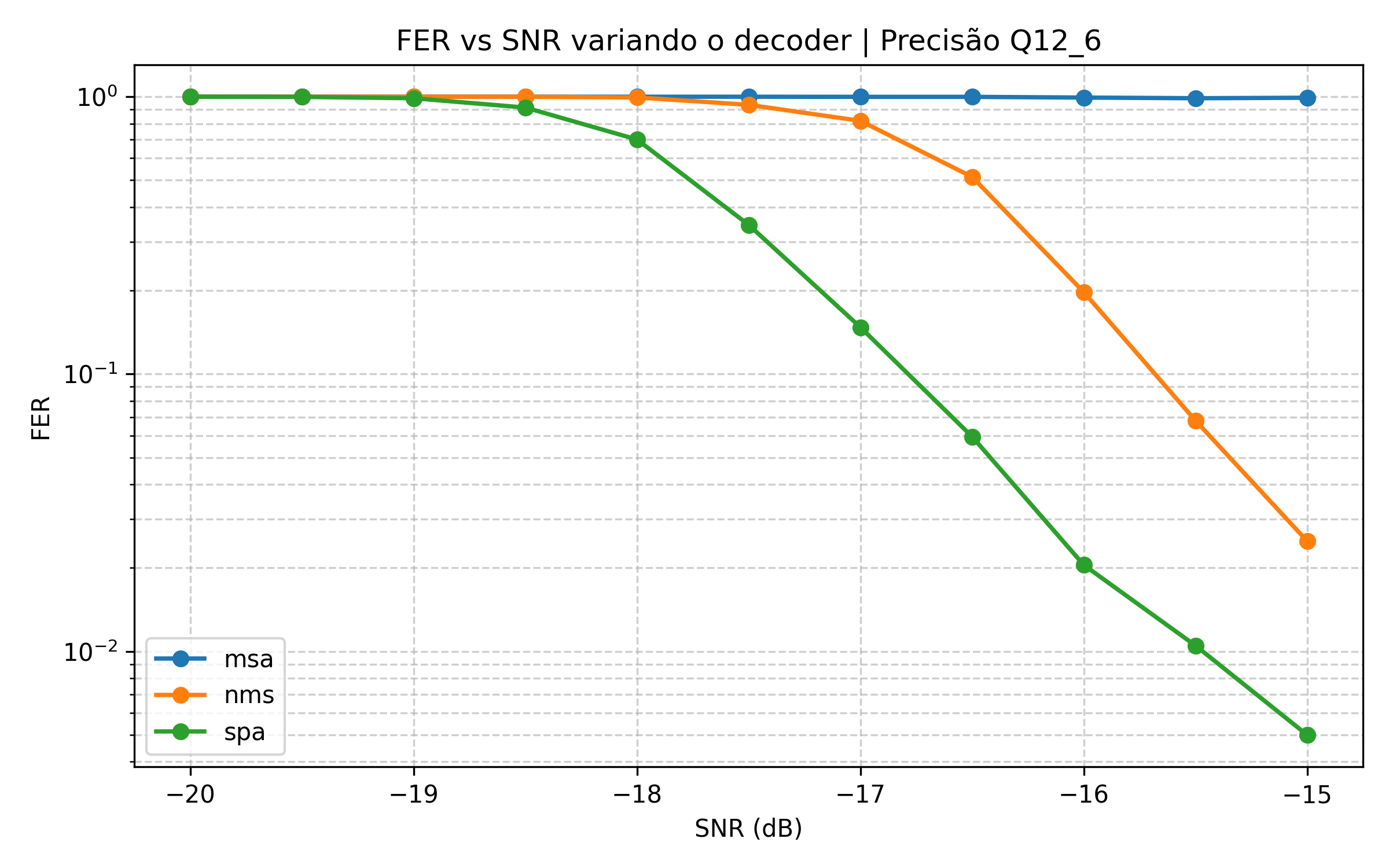}
    \caption{Bit error rate (FER) as a function of SNR for different decoders operating under the Q12.6 fixed-point format.}
    \label{fig: FER_Q12_6}
\end{figure}

Figure \ref{fig: FER_Q12_6} highlight the most representative FER comparison among the decoders when the fixed-point precision is kept constant.

The comparison indicates that SPA is the most reliable decoder among the evaluated strategies. For all precision settings, SPA exhibits the most pronounced waterfall region and the lowest FER over the relevant SNR interval, reinforcing its role as the strongest decoding reference in the present study. In contrast, the reduced-complexity decoders, especially MSA, show a stronger dependence on the adopted numerical format, which directly affects their ability to converge in the low-SNR regime.

\begin{figure}[H]
    \centering
    \includegraphics[width=1\linewidth]{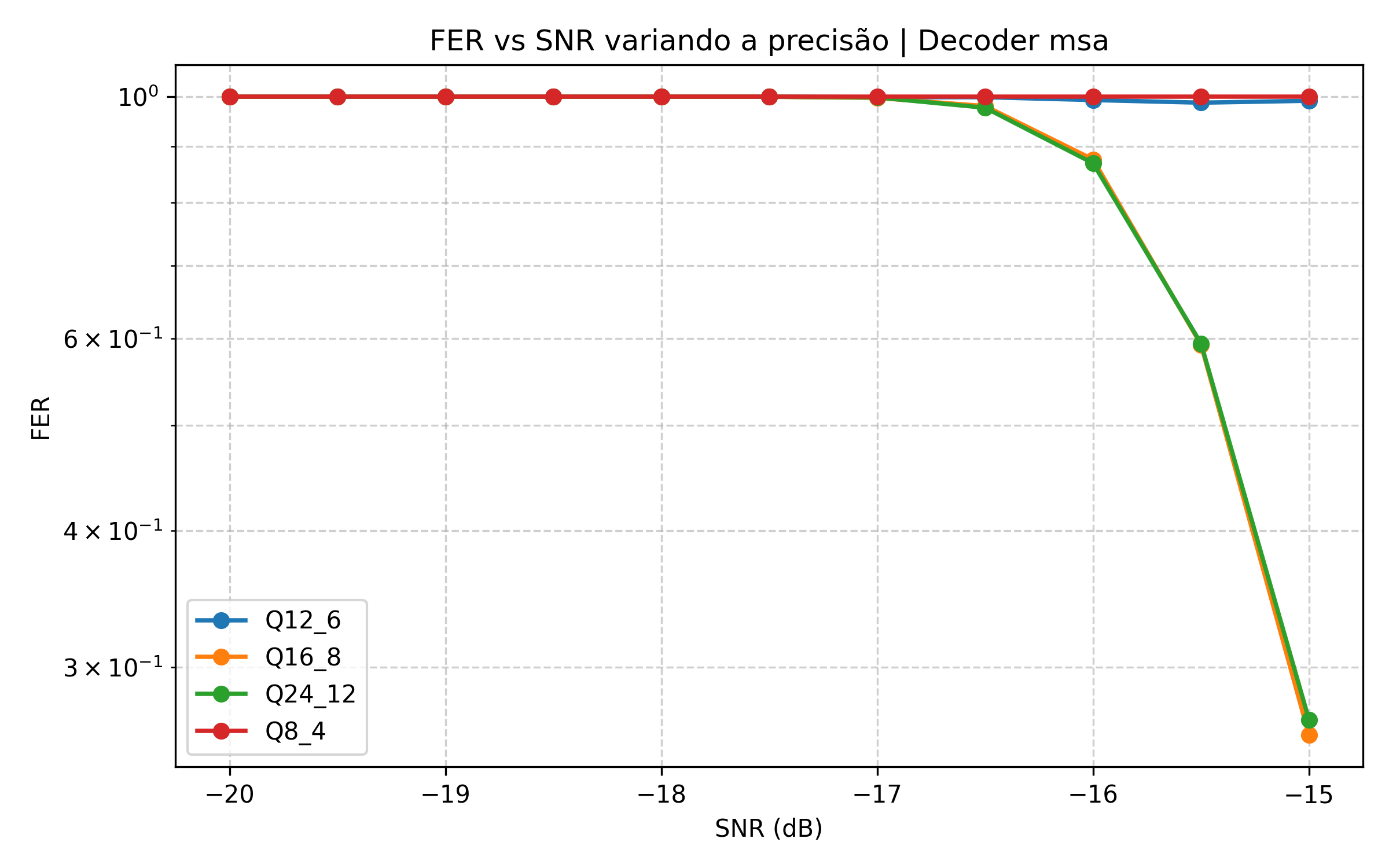}
    \caption{Bit error rate (FER) as a function of SNR for different fixed-point precision values using the MSA decoder.}
    \label{fig: FER_msa}
\end{figure}

\begin{figure}[H]
    \centering
    \includegraphics[width=1\linewidth]{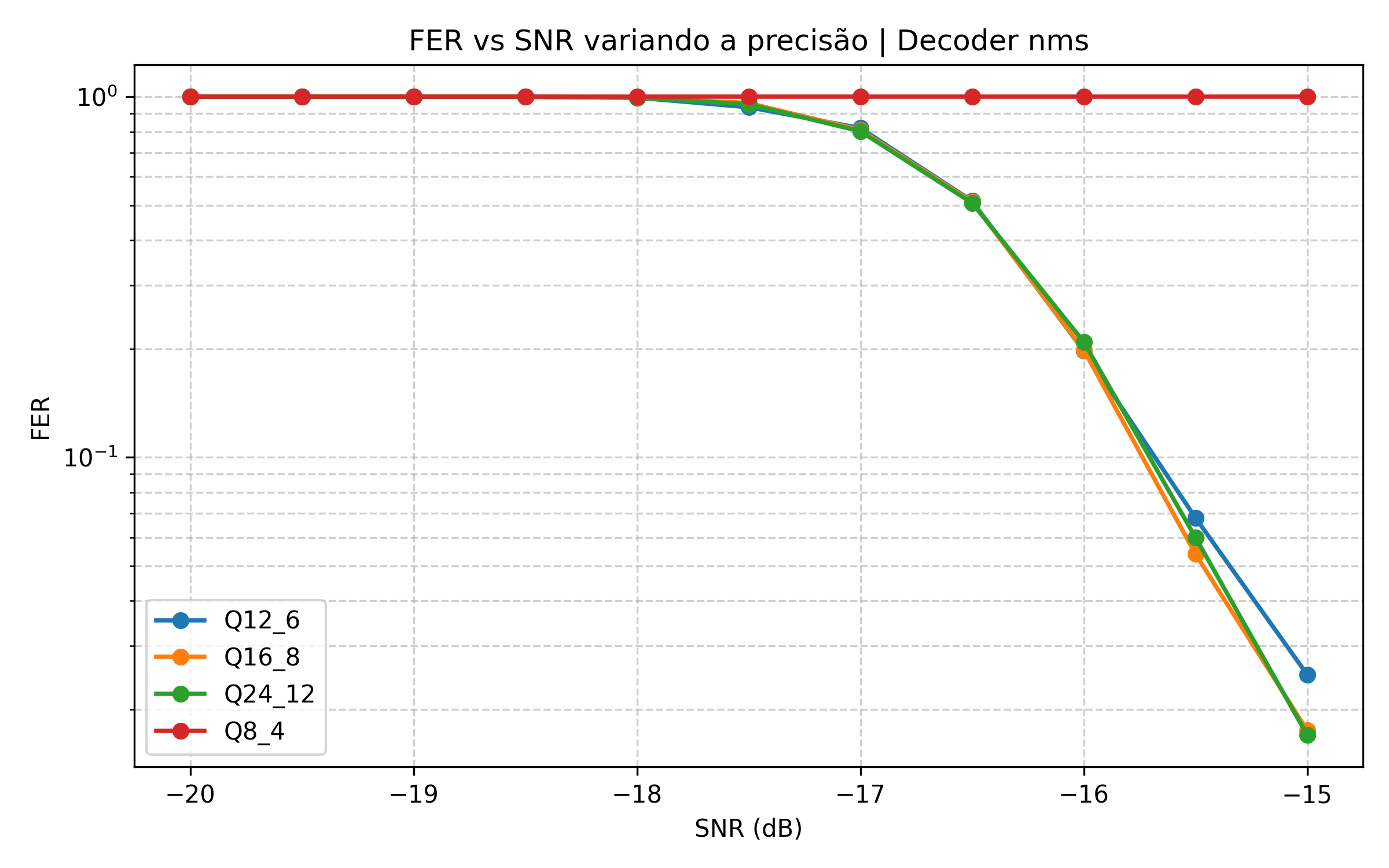}
    \caption{Bit error rate (FER) as a function of SNR for different fixed-point precision values using the NMS decoder.}
    \label{fig: FER_nms}
\end{figure}

\begin{figure}[!hbt]
    \centering
    \includegraphics[width=1\linewidth]{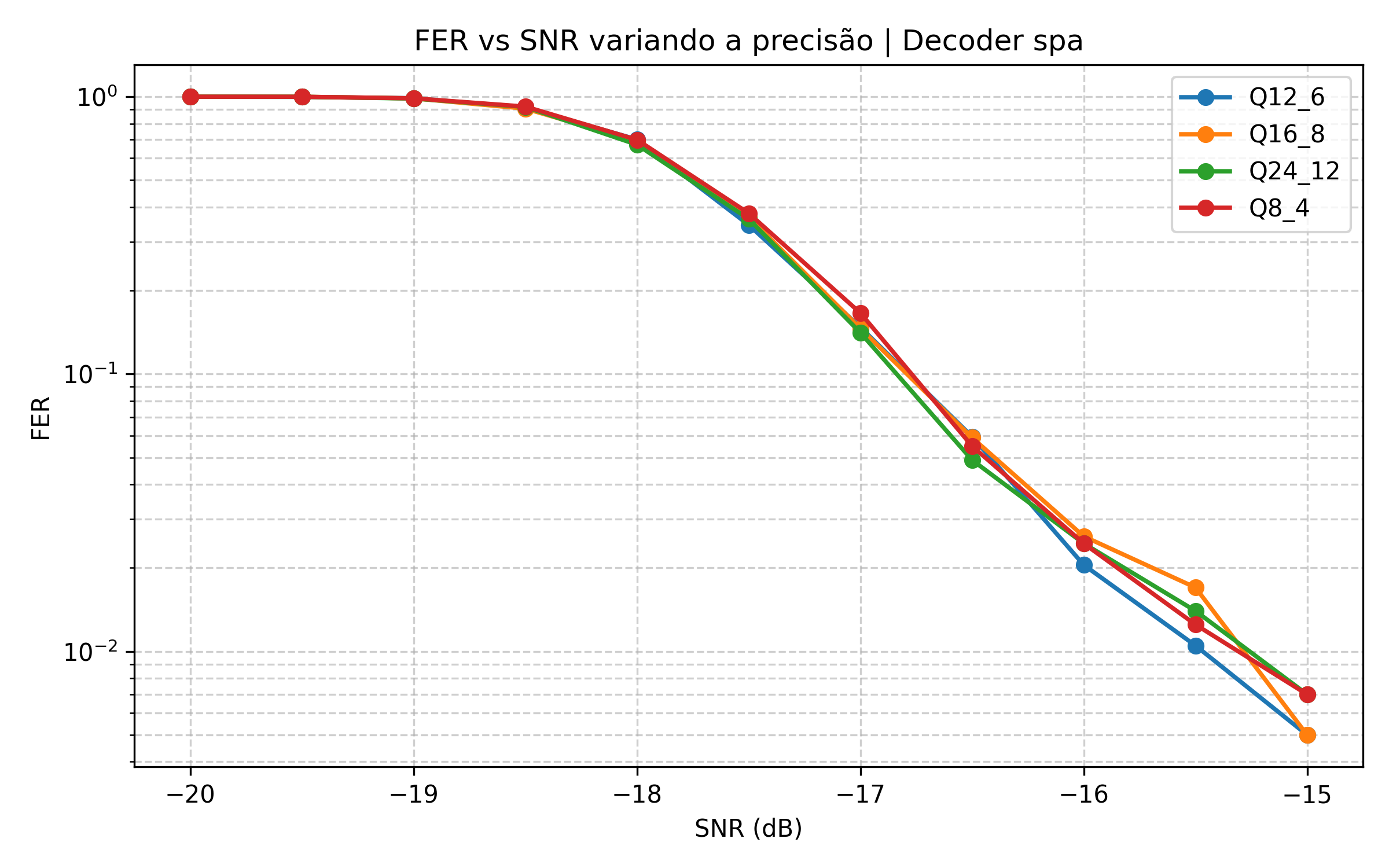}
    \caption{Bit error rate (FER) as a function of SNR for different fixed-point precision values using the SPA decoder.}
    \label{fig: FER_spa}
\end{figure}

Figures \ref{fig: FER_msa}, \ref{fig: FER_nms}, and \ref{fig: FER_spa} show the FER behavior of each decoder as a function of the tested fixed-point precision values.

The precision sweep reveals that both MSA and NMS achieve their most suitable operating condition under the Q16.8 and Q12.6 formats respectively. In this case, the decoders are able to preserve the expected improvement with increasing SNR while maintaining stable convergence. Lower precisions are progressively limited by quantization, which reduces the numerical fidelity of the exchanged messages and degrades decoding performance. On the other hand, larger formats do not provide further benefit and can become vulnerable to numerical saturation in intermediate operations, particularly when the increased fractional precision reduces the available dynamic range in fixed-width accumulators.

For SPA, the interpretation is slightly different. The decoder remains robust over a wider precision range, reaching its quantization-limited behavior at Q8.4 and extending up to Q24.12 without visible degradation in the final FER curves.

\section{Conclusion}

This work presented a comparative fixed-point evaluation of the SPA, MSA, and NMS decoders in a low-SNR LDPC decoding scenario, with emphasis on the influence of numerical precision on FER, and convergence behavior. The results show that decoder performance depends not only on the decoding rule itself, but also on how well the selected fixed-point format matches the numerical requirements of each algorithm.

Among the evaluated decoders, SPA delivered the best overall decoding results. It consistently achieved the lowest FER values, the clearest waterfall behavior, and the most stable convergence across the tested precision range. In addition, SPA remained effective even under reduced precision, which makes it particularly attractive from an implementation perspective.

For the reduced-complexity decoders, Q16.8 emerged as the optimal precision threshold, since it was the lowest tested format that still provided stable and reliable operation for all tested decoders algorithms. Under this condition, NMS clearly outperformed MSA and became the best simplified alternative to SPA. Lower precisions led to progressively stronger quantization effects.

From a practical standpoint, however, the most relevant outcome of this study is that the best overall scenario is obtained by combining SPA with the Q8.4 format. Although wider representations may still be numerically feasible for SPA, the Q8.4 format already preserves excellent decoding performance while significantly reducing the number of bits required in the fixed-point representation. This becomes especially important when the analysis is scaled to much larger parity-check matrices, where memory footprint, data movement, and arithmetic cost grow substantially. In this sense, SPA with Q8.4 provides the most favorable balance between decoding reliability and implementation efficiency within the evaluated conditions.

Overall, the results establish a clear ranking for the tested configurations: SPA is the most reliable decoder, NMS provides the best performance-complexity trade-off among the reduced-complexity alternatives, and MSA is the most sensitive to fixed-point constraints. Therefore, the selection of a fixed-point format must consider not only representational resolution, but also the numerical behavior of intermediate operations.

As a final outcome, this work provides practical guidance for hardware-oriented LDPC decoder design in the low-SNR regime. SPA should be regarded as the performance reference, while Q16.8 serves as the optimal threshold for reduced-complexity decoders. At the same time, SPA with Q8.4 stands out as the most attractive operating point when performance and scalability are considered jointly.

\section{Acknowledgements}
This work was fully funded by the project \textit{LDPC Code Design for Information Reconciliation in CV-QKD Optimized for Hardware Implementation}, supported by QuIIN – Quantum Industrial Innovation, the EMBRAPII CIMATEC Competence Center in Quantum Technologies. Financial resources were provided by the PPI IoT/Industry 4.0 program of the Brazilian Ministry of Science, Technology and Innovation (MCTI), under grant number 053/2023, in partnership with EMBRAPII.

\bibliographystyle{IEEEtran}
\bibliography{refs}

\end{document}